\documentclass[twocolumn,prb]{revtex4}
\usepackage{bm}

\begin{document}

\title{Dyakonov-Perel spin relaxation near metal-insulator transition and in hopping transport}

\author{B. I. Shklovskii}

\address{Department of Physics, University of Minnesota,
Minneapolis, MN 55455, USA}

\date{\today}

\begin{abstract}

In a heavily doped semiconductor with weak spin-orbital
interaction the Dyakonov-Perel spin relaxation rate is known to be
proportional to the Drude conductivity. We argue that in the case
of weak spin-orbital interaction this proportionality goes beyond
the Drude mechanism: at low temperatures it stays valid through
the metal-insulator transition and in the range of exponentially
small hopping conductivity.

\end{abstract}


\maketitle

Spin relaxation processes in semiconductors continue to attract
attention in connection with various spintronics
applications~\cite{KA,Dz,Fabian,Paul}. In crystals lacking a
center of inversion, for example in GaAs, spin of a free electron
experiences precession with the Larmor frequency ${ \bm \Omega}_{
\bm k}$, which is cubic in terms of components of the wave vector
$ {\bm k } $. Scattering of the electron randomly changes
direction of its wave vector $ \bm k $ and, therefore, the
direction of ${ \bm \Omega}_{ \bm k}$ leading to the angular
diffusion of spin magnetization ${ \bm S}$. This results in the
Dyakonov-Perel mechanism of spin relaxation ~\cite{DP}, which was
predicted 35 years ago and now is widely used to interpret spin
relaxation data in doped semiconductors~\cite{KA,Dz,Fabian,Paul}.
The spin relaxation time, $\tau_s$, is determined~\cite{DP} by
\begin{equation}
\tau_s^{-1} = \int_{0}^{\infty}<{\bm \Omega}_{ \bm k}(0){\bm
\Omega}_{\bm k}(t)>dt = {\Omega}^{2}\tau.~~\label{taus}
\end{equation}
Here ${\Omega}$ is the effective Larmor frequency averaged over
the electron energy distribution, $<{\bm \Omega}_{\bm k}(0){\bm
\Omega}_{\bm k}(t)>$ is the correlator of Larmor frequencies for
an electron at time difference $t$ and $\tau$ is the relaxation
time of the third order moment of the distribution function, which
we assume to be close to the electron momentum relaxation time.
Eq.~(\ref{taus}) is valid only for $\Omega \tau \ll 1$. The Drude
conductivity $\sigma = ne^2 \tau/m$, where $n$ is the
concentration of electrons, $e$ is the charge of an electron and
$m$ is its effective mass. This gives
\begin{equation}
\tau_s^{-1}= A \sigma, \label{prop}
\end{equation}
where $A \simeq {\Omega}^{2}m/ne^{2}$ is the dimensionless
coefficient.

The goal of this paper is to show that for a small enough
spin-orbit interaction Eq.~(\ref{prop}) is valid beyond the limits
of validity of the Drude mechanism of conduction. Let us imagine
that at a low temperature $T$ we vary the concentration of donors
$N$ in the semiconductor from $N \gg N_{c}$ to $N \ll N_{c}$,
where $N_c$ is the critical concentration of the metal-insulator
transition. Then at $N \gg N_c$ we deal with the Drude
conductivity and Eq.~(\ref{prop}) is valid. In the critical range
of the metal-insulator transition where $N > N_c$, but $N - N_c
\ll N_c$ the conductivity decreases as $e^2/\hbar \xi(N)$, where
the correlation length $\xi(N) = a[N_c/(N - N_{c})]^{\nu}$ and $a$
is the donor Bohr radius. This gives
\begin{equation}
\sigma \sim \frac{e^2}{\hbar a}[(N - N_c)/N_c]^{\nu} . \label{met}
\end{equation}
We argue below that for such "critical metal" Eq.~(\ref{prop}) is
still valid. At low temperatures one can define a narrow range $
\Delta N \ll N_c$, around $N_c$, where at $|N-N_c|< \Delta N$
metallic conductivity crosses over to the variable range hopping
conductivity (see calculation of $\Delta N$ below). Coulomb
interaction of electrons leads to the variable range hopping
following the Efros-Shklovskii (ES) law~\cite{ EfrSh}
\begin{equation}
\sigma(T) = \sigma_{0}\exp[-(T_0/T)^{1/2}], \label{hop}
\end{equation}
where $T_0 = Ce^2/\kappa(N)\xi(N)$, $C$ is the numerical
coefficient, $\xi(N)= a[N_c/(N_c - N)]^{\nu}$ is the localization
length and $\kappa(N)= \kappa [N_c/(N_c- N)]^{\zeta}$ is the
dielectric constant enhanced near the transition with respect of
its clean crystal value $\kappa$. We argue below that
Eq.~(\ref{prop}) is valid for the ES conductivity both near the
transition or in the lightly doped semiconductor, where $N <
N_{c}/2$, $\xi=a$ and $\kappa(N) = \kappa$.

Let us start from the metallic side of the transition, where the
conductivity of the critical metal is given by Eq.~(\ref{met}).
The reason of the conductivity reduction near the metal-insulator
transition is the interference leading to the non-Gaussian
diffusion. (One can say that electron dwells on some close loop
trajectories.) Still one can define electron trajectories, wave
vectors and velocities ${\bm v} = \hbar {\bm k} /m $ at each
trajectory (the dominating quadratic part of the electron spectrum
is isotropic). Then the conductivity is proportional to the
diffusion coefficient
\begin{equation}
D =\int_{0}^{\infty}<{\bm v(0)}{\bm v(t)}> dt, \label{dif}
\end{equation}
where $<{\bm v(0)}{\bm v(t)}>$ is the correlator of electron
velocities. On the other hand, one can write a scaling estimate
\begin{equation}
\tau_s^{-1}=\int_{0}^{\infty}<{\bm \Omega}_{\bm k}(0){\bm
\Omega}_{\bm k}(t)>dt \sim {\Omega}^{2}\tau \sim D
\frac{\Omega^2}{<v^{2}>}. \label{omega}
\end{equation}
This proves Eq.~(\ref{prop}) for the critical metal case.

Let us now consider the Dyakonov-Perel mechanism of spin
relaxation for the hopping conductivity. Any hopping transport can
be considered as a result of fast tunnel hops from one localized
state to another alternating with exponentially long waiting
periods in each localized state. While waiting an electron has $k
= v = 0$ and, therefore, is not relaxing its spin via the
Dyakonov-Perel mechanism. On the other hand, an electron tunneling
between two localized states has the real trajectory and the real
displacement, which it traverses in imaginary time and, therefore,
it has the imaginary $k$ and $v$. Therefore, its spin experiences
precession in the course of tunneling. Its Larmor frequency
$\Omega \propto k^3$ is imaginary, too. But because time is
imaginary the angle of rotation in the course of the hop is real.
This real angle is proportional to the real displacement of the
electron and the direction of rotation is related to the direction
of the hop.

The fraction of time during which the electron hops or, in other
word, tunnels is proportional to $\exp[-(T_0/T)^{1/2}]$. This is
why the hopping conductivity has this small exponential factor.
But $\tau_s^{-1}$ should have the same small factor because as we
explained relaxation happens only during hops. It is clear,
therefore, Eq.~(\ref{prop}) should be valid for the ES law, at
least in the exponential sense.

One can improve these arguments using the language of redefined
correlators $<{\bm \Omega}_{\bm
 k}^{*}(0){\bm \Omega}_{k}^{*}(t)>$ and $<{\bm v^{*}(0)}{\bm v^{*}(t)}>$.
In this correlators ${\bm \Omega}_{k}^{*}(t)$  and ${\bm
v}^{*}(t)$ are the rotation angle during a hop and the hop
displacement divided by the waiting time, respectfully. These
correlators now decay on exponentially large times because all the
waiting times are included in their definition. In the hopping
conductivity regime the first correlator is responsible for the
spin relaxation rate $\tau_s^{-1}$, while the second one
calculated for a long enough time history of an electron is
related to the diffusion coefficient and the conductivity. These
correlators are obviously proportional to each other, what again
leads to Eq.~(\ref{prop}). Note that our approach to spin
relaxation in a lightly doped semiconductor is completely
different from the one suggested by Kavokin~\cite{Kavokin} and
based on the role of the anisotropic exchange between electrons
localized on different donors. While we are talking about
Dyakonov-Perel relaxation related to a single electron diffusion
in space, Kavokin relies on rotation of spin of a localized
electron in the collective field of other localized electrons.

Let us make a comment about the range of concentrations, where
crossover between Eq.~(\ref{met}) and Eq.~(\ref{hop}) takes place,
while staying away from any discussion of the mechanism of
conductivity in this range. At low temperature the relative width
of this range is small, $\Delta N/N_{c}<<1$. Indeed, one can
estimate $\Delta N$ equating $T_0(N)$ to $T$ and identifying
$\Delta N$ with $N_c - N$. This gives $\Delta N/N_c
=[T/(e^2/\kappa a)]^{1/(\nu + \zeta)}$. It is known from
experiments~\cite{ EfrSh} that $\nu + \zeta \simeq 2$. As we
argued above Eq.~(\ref{prop}) is valid on both sides of the
crossover range $\Delta N$. This means that Eq.~(\ref{prop}) is
valid in the crossover range as well.

Above we have concentrated on the three-dimensional case. In two
dimensions validity of Eq.~(\ref{prop}) for the hopping
conductivity can be demonstrated even more transparently. Let us
consider the 2DEG without structural inversion asymmetry in the
$(001)$-plane of GaAs crystal and assume that initially electron
spins are polarized along $z$-axis perpendicular to 2DEG plane.
Then at times smaller than $\tau_{s}$ the spin magnetization $\bm
S$ evolves following to the equation
\begin{equation}
dS_{x}/dt = \Omega_{y}S_{z},~~~ dS_{y}/dt = - \Omega_{x}S_{z},
\label{evolution}
\end{equation}
where
\begin{equation}
\Omega_{x}= \gamma k_{y} (k_{x}^{2}-k_{z}^{2}),~~~\Omega_{y}=
-\gamma k_x(k_{y}^{2}-k_{z}^{2}). \label{omega}
\end{equation}

For a narrow quantum well the momentum components $k_{x}^{2},
k_{y}^{2}$ are much smaller than $k_{z}^{2}$ and, therefore, can
be neglected in the right sides of Eqs. ~(\ref{omega})~\cite{dk}.
Then one can easily calculate the change of the spin magnetization
$\delta {\bm S}$ during the time $\delta t \ll \tau_s$. This gives
\begin{equation}
\delta {\bm S}/S_z = \gamma (m/\hbar){\bm v} \delta t,
\label{angledisplacement}
\end{equation}
i.e. the angle of rotation of the spin magnetization is
proportional to the electron displacement in the plane of quantum
well. This leads directly to Eq.~(\ref{prop}), both for the case
of metallic conductivity and for the hopping transport. While in
the in the latter one both ${\bm v}$ and $t$ are imaginary
quantities, the angle of rotation of the spin magnetization and
the electron displacement are real and as we see initial
rotational diffusion of ${\bm S}$ and diffusion in the real space
are related as tightly as for the metallic conduction.

This means that in the range of the ES variable range hopping both
in three and two dimensions the Dyakonov-Perel spin relaxation
rate is very small and exponentially decreases with temperature.
\begin{equation}
\tau_s^{-1} \propto \exp[-(T_0/T)^{1/2}]. \label{Tdependence}
\end{equation}
As function of donor concentration $N$ the rate has to
exponentially decrease with growth of $T_0$, while $N$ is still in
the critical range of transition $N_c - N \ll N_c$. At $N <
N_{c}/2$ the temperature $T_0$ saturates at $T_{0} = Ce^2/a\kappa$
and $\tau_s^{-1}$ saturates at very small level exponentially
dependent on $T$.

Of course, other mechanisms of spin relaxation can take over at
weak doping and at low temperature~\cite{Dz,KA,Fabian}, but
because Dyakonov-Perel relaxation typically is the dominating
mechanism this crossover may happen only at a very small
relaxation rates.

Let us make a comment about measurement of $\tau_s$ in the hopping
regime. In a typical experiment polarized electrons are created in
the conduction band and may experience few scattering events
before being captured by donors with characteristic capture time
$\tau_c$. Thus, they may loose a fraction of their polarization by
with DP spin relaxation time of free electrons $\tau_f$. If
$\tau_c < \tau_f$ they get captured before loosing spin in the
conduction band. Then $\tau_s$ calculated above describes
relaxation of practically all the polarization. In the opposite
case, when $\tau_c > \tau_f$ only a small fraction of polarization
of the order of $\tau_f/\tau_c$ relaxes via hopping, while
majority of the polarization relaxes faster.

In a pump-probe experiments~\cite{KA} this means that hopping
relaxation dominates only at times larger than $\tau_f \ln
\tau_c/\tau_f$. Thus, in this case, hopping $\tau_s$ describes the
tail of the spin relaxation. On the other hand, in continuous wave
excitation experiments spin relaxation also happens first in the
conduction band and then via hopping on donors. For $\tau_c <
\tau_f$ a standard way~\cite{Dz} to measure $\tau_s$ directly
leads to the hopping spin relaxation time. On the other hand,
measuring hopping spin relaxation time by this method at $\tau_c >
\tau_f$ is difficult and one needs more delicate methods like
direct optical readout of donor spins.

In $n$-type GaAs the dependence of low temperature spin relaxation
on doping level was recently studied~\cite{Dz}. It was
interpreted~\cite{Dz} with the help of the mechanism of
anisotropic exchange (immediately below the metal insulator
transition) and by the hyperfine interaction with nuclei (at very
small doping). These data look as if there is no substantial range
of doping, where hopping Dyakonov-Perel relaxation dominates and
$\tau_s^{-1}$  decreases with the concentration of donors
proportionally to the hopping conductivity. This could be a result
of the discussed above masking effect of spin loss during energy
relaxation in the conduction band. If this is true, a pump-probe
experiment should reveal the DP hopping relaxation in the long
time tail of relaxation.

I am grateful to P. A. Crowell, A. P. Dmitriev,  M. I. Dyakonov,
Al. L. Efros, J. Fabian, V. Yu. Kachorovskii, A. I. Larkin, I. S.
Lyubinskiy and V. I. Perel for useful discussions. I acknowledge
hospitality of Aspen Center for Physics, where this paper was
written.

\end{document}